\documentclass[prc,showpacs,showkeys,superscriptaddress,nofootinbib,twocolumn]{revtex4}
\usepackage{graphicx,color,epstopdf,mathrsfs,amsmath,amssymb,amsthm,amsopn,mathbbol,bm} 
\usepackage{feynmp}

\def\scr#1{\mbox{\scriptsize #1}}
\newcommand{\Tr}{{\mathrm{Tr}}}
\def\ii{\mathrm{i}}

\def\lsim{\lesssim}
\def\gsim{\gtrsim}

\begin{document}
\title{Renormalization of a gapless Hartree-Fock approximation
to a theory with spontaneously broken $O(N)$-symmetry
} 

\author{Yu. B. Ivanov}\thanks{e-mail: Y.Ivanov@gsi.de}
\affiliation{Gesellschaft
 f\"ur Schwerionenforschung mbH, Planckstr. 1,
D-64291 Darmstadt, Germany}
\affiliation{Kurchatov Institute, Kurchatov
sq. 1, Moscow 123182, Russia}
\author{F. Riek}\thanks{e-mail: F.Riek@gsi.de}
\affiliation{Gesellschaft
 f\"ur Schwerionenforschung mbH, Planckstr. 1,
D-64291 Darmstadt, Germany}
\author{H. van Hees}\thanks{e-mail: hees@comp.tamu.edu}
\affiliation{Cyclotron Institute, Texas A\&M University, 
College Station, Texas 77843-3366, USA}
\author{J. Knoll}\thanks{e-mail: J.Knoll@gsi.de}
\affiliation{Gesellschaft
 f\"ur Schwerionenforschung mbH, Planckstr. 1, 
D-64291 Darmstadt, Germany}

\begin{abstract}
  The renormalization of a gapless $\Phi$-derivable Hartree--Fock
  approximation to the $O(N)$-symmetric $\lambda\phi^4$ theory is
  considered in the spontaneously broken phase. This kind of approach was
  proposed by three of us in a previous paper \cite{IRK05} in order to preserve all the
  desirable features of $\Phi$-derivable Dyson-Schwinger resummation
  schemes (i.e., validity of conservation laws and thermodynamic
  consistency) while simultaneously restoring the Nambu--Goldstone theorem
  in the broken phase. It is shown that unlike for the conventional
  Hartree--Fock approximation this approach allows for a scale-independent
  renormalization in the vacuum. However, the scale dependence still
  persists at finite temperatures. Various branches of the solution are
  studied. The occurrence of a limiting temperature inherent in the
  renormalized Hartree--Fock approximation at fixed renormalization
  scale $\mu$ is discussed.
\end{abstract}

\date{\today}
\maketitle

\begin{fmffile}{phi-func}
\def\fmfsdot#1{\fmfv{decor.shape=circle,decor.filled=full,decor.size=1.4thick}
{#1}}
\def\fmffdot#1{\fmfv{decor.shape=circle,decor.filled=full,decor.size=2.4thick}
{#1}}
\def\fmfcross#1#2{
\fmfv{decor.shape=cross,decor.angle=#1,decor.size=5thick}{#2}
}
\def\PhiF{
\parbox{14mm}{\centerline{
\begin{fmfgraph*}(10,10)
\fmfpen{thick}
\fmfleft{l}
\fmfright{r}
\fmftop{t}
\fmfbottom{b}
\fmfforce{(0.5w,0.5h)}{m}
\fmf{plain}{t,b}
\fmf{plain}{l,r}
\fmfsdot{m}
\fmfcross{0}{b,l,r,t}
\end{fmfgraph*}}
}}
\def\JPhiF{
\parbox{14mm}{\centerline{
\begin{fmfgraph*}(10,10)
\fmfpen{thick}
\fmfleft{l}
\fmfright{r}
\fmftop{t}
\fmfbottom{b}
\fmfforce{(0.5w,0.5h)}{m}
\fmf{plain}{t,b}
\fmf{phantom}{l,ll}
\fmf{plain}{m,r}
\fmf{plain,width=1}{ll,m}
\fmffdot{m}
\fmfcross{0}{b,r,t}
\end{fmfgraph*}}
}}
\def\PhiTtad{
\parbox{12mm}{\centerline{
\begin{fmfgraph*}(10,10)
\fmfpen{thick}
\fmfleft{lb,lt}
\fmfforce{0.5w,1.3h}{t}
\fmfright{rb,rt}
\fmfforce{(0.5w,0.5h)}{m}
\fmf{plain}{lb,m,rb}
\fmf{plain,right=1}{m,t,m}
\fmfsdot{m}
\fmfcross{45}{lb,rb}
\end{fmfgraph*}}
}}
\def\JPhiTtad{
\parbox{12mm}{\centerline{
\begin{fmfgraph*}(10,10)
\fmfpen{thick}
\fmfleft{lb,lt}
\fmfforce{0.5w,1.3h}{t}
\fmfright{rb,rt}
\fmfforce{(0.5w,0.5h)}{m}
\fmf{phantom}{lb,llb}
\fmf{plain,width=1}{llb,m}
\fmf{plain}{m,rb}
\fmf{plain,right=1}{m,t,m}
\fmffdot{m}
\fmfcross{45}{rb}
\end{fmfgraph*}}
}}
\def\SPhiTtad{
\parbox{12mm}{\centerline{
\begin{fmfgraph*}(10,10)
\fmfpen{thick}
\fmfleft{lb,l,lt}
\fmfforce{0.5w,1.3h}{t}
\fmfright{rb,r,rt}
\fmfforce{(0.5w,0.5h)}{m}
\fmf{plain,width=1}{ll,m,rr}
\fmf{plain}{lb,m,rb}
\fmf{phantom,tension=2}{l,ll}
\fmf{phantom,tension=2}{r,rr}
\fmffdot{m}
\fmfcross{45}{lb,rb}
\end{fmfgraph*}}
}}
\def\Phieight{
\parbox{12mm}{\centerline{
\begin{fmfgraph*}(10,16)
\fmfpen{thick}
\fmfleft{lt,lb}
\fmfforce{0.5w,0h}{b}
\fmfforce{0.5w,1h}{t}
\fmfright{rt,rb}
\fmfforce{(0.5w,0.5h)}{m}
\fmf{plain,right=1}{m,t,m}
\fmf{plain,right=1}{m,b,m}
\fmfsdot{m}
\end{fmfgraph*}}
}}
\def\Phisun{
\parbox{25mm}{\centerline{
\begin{fmfgraph*}(20,10)
\fmfpen{thick}
\fmfleft{l}
\fmfforce{0.2w,0.5h}{ll}
\fmfforce{0.8w,0.5h}{rr}
\fmfright{r}
\fmf{plain}{l,ll,rr,r}
\fmf{plain,right=0.75}{ll,rr,ll}
\fmfsdot{ll,rr}
\fmfcross{0}{l,r}
\end{fmfgraph*}}
}}
\def\Phifootball{
\parbox{18.5mm}{\centerline{
\begin{fmfgraph*}(16,25)
\fmfpen{thick}
\fmfleft{l}
\fmfright{r}
\fmf{plain,right=0.34}{l,r,l}
\fmf{plain,right=0.8}{l,r,l}
\fmfsdot{l,r}
\end{fmfgraph*}}
}}
\def\SPhieight{
\parbox{12mm}{\centerline{
\begin{fmfgraph*}(10,16)
\fmfpen{thick}
\fmfleft{lt,l,lb}
\fmfforce{0.5w,0h}{b}
\fmfforce{0.5w,1h}{t}
\fmfright{rt,r,rb}
\fmfforce{(0.5w,0.5h)}{m}
\fmf{plain,right=1}{m,t,m}
\fmf{plain,width=1}{ll,m,rr}
\fmf{phantom,tension=2}{l,ll}
\fmf{phantom,tension=2}{r,rr}
\fmffdot{m}
\end{fmfgraph*}}
}}
\def\JPhisun{
\parbox{25mm}{\centerline{
\begin{fmfgraph*}(20,10)
\fmfpen{thick}
\fmfleft{l}
\fmfforce{0.2w,0.5h}{ll}
\fmfforce{0.8w,0.5h}{rr}
\fmfright{r}
\fmf{plain,width=1}{l,ll}\fmf{plain}{ll,r,rr}
\fmf{plain,right=0.75}{ll,rr,ll}
\fmffdot{ll}
\fmfsdot{rr}
\fmfcross{0}{r}
\end{fmfgraph*}}
}}

\section{Introduction}

Self-consistent $\Phi$-derivable approximations were introduced long ago in
the context of the nonrelativistic many-body problem~\cite{Luttinger,Baym}
and then extended to relativistic quantum field
theory~\cite{Cornwall,Baym-Grin}. Recently the interest in this method has
been revived in view of its fruitful applications to calculations of the
thermodynamic properties of the quark--gluon plasma~\cite{Blaizot99} and to
non-equilibrium quantum-field dynamics~\cite{IKV98a,Berges01,Cooper03}, in
particular in terms of the off-shell kinetic equation~\cite{IKV99}.

$\Phi$-derivable approximations are preferable for the dynamical treatment
of a system, since they fulfill the conservation laws of energy, momentum,
and charge~\cite{Baym,IKV98a,IKV99}.  Moreover, the $\Phi$-derivable scheme
also guarantees the thermodynamic consistency of an
approximation~\cite{Baym}, which makes it advantageous also for
thermodynamic calculations. However, the $\Phi$-derivable scheme has its
generic problems which were also realized long ago~\cite{Baym-Grin,HM65}.

The first problem is related to the fact that $\Phi$-derivable
Dyson-Schwinger resummation schemes violate Ward--Takahashi identities
beyond the one-point level. This, in particular, results in the violation
of the Nambu--Goldstone (NG) theorem~\cite{Baym-Grin,HM65,Aouissat,HK3} in
the phase of spontaneously broken symmetry. On the other hand, so called
``gapless'' approximations~\cite{HM65} respect the NG theorem (which is
referred to as the Hugenholtz--Pines theorem in physics of Bose--Einstein
condensed systems), though violate conservation laws and thermodynamic
consistency. In Ref.~\cite{HK3} it was shown that any $\Phi$-derivable
approximation can be corrected in such a way that it respects the NG
theorem and becomes gapless. However, such modifications again violate
conservation laws and thermodynamic consistency and, hence, leads back to
the problems of the gapless scheme. Recently three of us have proposed a
phenomenological way to construct a ``gapless $\Phi$-derivable''
Hartree--Fock (gHF) approximation to the $\lambda \phi^4$ theory in the
phase of spontaneously broken $O(N)$ symmetry~\cite{IRK05}.  This
approximation simultaneously preserves all the desirable features of
$\Phi$-derivable schemes and respects the Nambu--Goldstone theorem in the
broken phase.  The treatment of Ref.~\cite{IRK05} was based on a naive
renormalization, where all divergent terms were simply omitted. This was
done in order to avoid possible confusions between effects of restoring the
NG theorem and those related to renormalization. In the present paper we
return to the issue of renormalization.

The renormalization of the $\Phi$-derivable approximations is precisely the
second main problem. Following Baym and Grinstein~\cite{Baym-Grin} it was
believed that renormalization of $\Phi$-derivable approximations is
possible only with medium (e.g., temperature) dependent counter terms,
which is inconsistent with the goal of renormalization.  Great progress in
the proper renormalization of such schemes was recently achieved in
Refs.~\cite{HK3,HK1,Reinosa1,Berges04,Cooper05}.  As the main result it was
shown that partial resummation schemes can indeed be renormalized with
medium-independent counter terms provided the scheme is generated from a
two-particle irreducible (2PI) functional, i.e., a $\Phi$-functional.
Still, as we are going to demonstrate below, certain problems remain in the
case of spontaneously broken symmetry.

As an example case we investigate the $O(N)$ model in the spontaneously
broken phase which is a traditional touchstone for new theoretical
approaches, well applied to a variety of physical phenomena, such as the
chiral phase transition in nuclear matter.  Thus we continue the discussion
of the ``gapless Hartree-Fock approximation started in the previous
paper~\cite{IRK05} and investigate its features towards renormalization in
comparison to the standard HF-approximation.

\section{Gapless Hartree--Fock (${\mbox{\bf gHF}}$) Approximation}

We consider the $O(N)$-model Lagrangian 
\begin{eqnarray}
\begin{split}
{\cal L} =& \frac{1}{2} (\partial_\mu \phi_a)^2 - \frac{1}{2} m^2 \phi^2
- \frac{\lambda}{4N} (\phi^2)^2 \\&+ H\cdot \phi, 
\end{split}
\end{eqnarray}
where $\phi = (\phi_1,\phi_2,...,\phi_N)$ is an $N$-component scalar
field, $\phi^2 = \phi_a \phi_a$, with summation over $a$ implied. For
$H=0$ this Lagrangian is invariant under $O(N)$ rotations of the
fields. If $H=0$ and $m^2 < 0$, the symmetry of the ground state is
spontaneously broken 
down to $O(N-1)$, with $N-1$ Goldstone bosons (pions). The external
field $H\cdot \phi = H_a \phi_a$ is a term which explicitly breaks the
$O(N)$ symmetry.  It is introduced to give the physical
value of $140$~MeV to the pion mass.
 
The effective action $\Gamma$ for this Lagrangian in the real-time
formalism is defined as (cf. Ref. \cite{IKV98a})
\begin{eqnarray}
\begin{split}
\label{Gamma-Phi}
\Gamma\{\phi,G\}&=I_0(\phi)+
\frac{\ii}{2}\Tr\left(\ln G^{-1}\right)\\
&+\frac{\ii}{2}\Tr\left(D^{-1}G-1\right)+\Phi_{\scr{real-time}}\{\phi,G\}, 
\end{split}
\end{eqnarray}
where $\phi$ is the expectation value of the field, $G$ is the Green's
function, $D$ is the \emph{free} Green's function, $I_0(\phi)$ is the
\emph{free} classical action of the $\phi$ field, $\Tr$ implies space--time
integration and summation over field indices $a,b,...$ All the
considerations below are performed in terms of the thermodynamic $\Phi$
functional which differs from $\Phi_{\scr{real-time}}$ in the factor of
$\ii\beta$, where $\beta=1/T$ is the inverse temperature. In the case of a
spatially homogeneous thermodynamic system, an additional factor appears:
the volume $V$ of the system. Thus, the thermodynamic $\Phi$ is
$$\Phi=(-\ii   T/V)\Phi_{\scr{real-time}}.$$ 
In terms of the CJT formalism~\cite{Cornwall}, the same effective
action $\Gamma$ is given as (e.g., cf. Ref. \cite{Berges01})
\begin{eqnarray}
\begin{split}
\label{Gamma-2}
\Gamma\{\phi,G\}&=I(\phi)+
\frac{\ii}{2}\Tr\left(\ln G^{-1}\right)\\
&+\frac{\ii}{2}\Tr\left(D^{-1}_\phi G-1\right)+\Gamma_2\{\phi,G\}, 
\end{split}
\end{eqnarray}
i.e., already in terms of the \emph{tree-level} Green's function,
$D_\phi(1,2)=\delta^2 I/\delta\phi_2\delta\phi_1$, and the \emph{full}
classical action of the $\phi$ field, $I(\phi)$.  In the thermodynamic
limit, the effective potential, $V_{\scr{CJT}}$, and its interaction part,
$V_2$, are defined as
$$V_{\scr{CJT}}=(-\ii T/V)\Gamma, \quad V_2=(-\ii T/V)\Gamma_2.$$ 
Naturally,  $V_2$ is similar to $\Phi$ but is not quite the same. 
Contrary to $V_2$, 
the $\Phi$ functional includes all 2PI
  interaction terms, i.e. also those of zero and first loop order
  which result from interactions with the classical field (first two
  graphs in (\ref{Phi-Hartree})).
  
  The gHF approximation to the $O(N)$ theory is defined by the $\Phi$
  functional \cite{IRK05}
\begin{eqnarray}
\label{Phi-Hartree}
\Phi_{\scr{gHF}}=\PhiF
+\hspace*{-2mm}\PhiTtad\hspace*{-2mm}
+\hspace*{-2mm}\Phieight\hspace*{-2mm}+\Delta\Phi, 
\end{eqnarray}
where the diagrams on the r.h.s. constitutes the conventional HF
approximation, while \emph{the phenomenological NG-theorem-restoring
  correction $\Delta\Phi$} is specified below, see Eq. (\ref{dPhi-ab}).
Here the crossed pins denote the classical fields $\phi_a$, and loops are
tadpoles
\begin{eqnarray}
\label{Pab}
\SPhieight = Q_{ab} = \int_\beta d^4 k G_{ab}(k)
\end{eqnarray}
in terms of $G_{ab}$ Green's functions, where the Matsubara summation
\begin{eqnarray}
\label{int-T}
\int_\beta d^4 q f(q) 
\equiv T \sum_{n=-\infty}^\infty 
\int \frac{d^3 q}{(2\pi)^3} f(2\pi\ii nT,\vec{q}) 
\end{eqnarray}
is implied with $T$ being a finite temperature. 

Within the $\Phi$-derivable scheme the r.h.s. of the equations of motion
for the classical field ($J$) and the Green's function (self-energy $\Sigma$)
follow from the functional variation of $\Phi_{\scr{gHF}}$ with respect to
the classical field $\phi$ and Green's function $G$, respectively
\begin{eqnarray}
\label{Phi-J}
\Box\phi+m^2\phi&=&J=\displaystyle 
\frac{\delta\Phi_{\scr{gHF}}}{\delta\phi}
\cr
&=&\JPhiF+\JPhiTtad,
\\[2mm]
\label{Phi-S}
G^{-1}-D^{-1}&=&
\Sigma=\displaystyle 
2 \frac{\delta\Phi_{\scr{gHF}}}{\delta G}
\cr
&=&\hspace*{-1mm}\SPhiTtad\hspace*{-1mm}
+\hspace*{-1mm}\SPhieight\hspace*{-1mm}
+2 \frac{\delta\Delta\Phi}{\delta G}, 
\end{eqnarray}
where $D$ is the free propagator. 

The $\Delta\Phi$ correction, introduced in Ref. \cite{IRK05}, is
unambiguously determined proceeding from the following requirements: (i) it
restores the NG theorem in the broken-symmetry phase, (ii) it does not
change results in the phase of restored $O(N)$ symmetry, because there is
no need for it, (iii) it does not change the HF equation for the classical
field, since the conventional $\Phi$-derivable and gapless schemes
\cite{Baym-Grin,HM65} provide the same classical-field equation already
without any modifications. In particular, due to this latter requirement
the $\Delta\Phi$ correction does not contribute to the classical-field
equation (\ref{Phi-J}). Proceeding from these requirements, this
$\Delta\Phi$ can be presented in manifestly $O(N)$ symmetric form
\begin{eqnarray}
\label{dPhi-ab}
\Delta\Phi = -
\frac{\lambda}{2N} \left[N(Q_{ab})^2 - (Q_{aa})^2\right].
\end{eqnarray}
Here and below, summation over repeated indices
$a,b,c,...$ is implied, if it is not pointed out otherwise.  

The nature of this correction can be understood as follows. For the full
theory, i.e., when all diagrams in the $\Phi$ functional are taken into
account, the gapless and $\Phi$-derivable schemes are identical and both
respect the NG theorem. The conventional $\Phi$-derivable HF approximation
omits an infinite set of diagrams which is necessary to restore its
equivalence with the gapless scheme.  The $\Delta\Phi$ correction to the HF
approximation takes into account a part of those omitted diagrams (at the
level of the actual approximation), and thus restores this
equivalence in the pion sector.  For the further discussion we switch to
the notation in terms of the CJT effective potential, see
e.g.~\cite{Cornwall,Lenaghan,Nemoto}, in order to comply with previous
considerations in the literature.

The manifestly symmetric form of the CJT effective potential in the gHF
approximation reads
\begin{widetext}
\begin{eqnarray}
\label{V-H}
V_{\scr{gHF}}(\phi,G) &=& \frac{1}{2} m^2 \phi^2 + 
\frac{\lambda}{4N} (\phi^2)^2 - H\cdot \phi +
\frac{1}{2} \int_\beta d^4 k \ln\det G^{-1}(k)
\cr
&+&
\frac{1}{2} \int_\beta d^4 k \left\{
\left[\left(k^2+m^2\right)\delta_{ab} + \frac{\lambda}{N} 
\left(\phi^2 \delta_{ab} + 2 \phi_a\phi_b\right)
\right] G_{ba}(k)-1\right\}
\cr
&+&
\frac{\lambda}{4N} 
\left(Q_{aa} Q_{bb} + 2 Q_{ab}Q_{ba} \right)+\Delta\Phi,
\end{eqnarray}
\end{widetext}
e.g., cf. \cite{Nemoto}.  All
quantities are symmetric with respect to permutations of indices.

Since the self-energy (\ref{Phi-S}) is momentum independent, the
general form of the Green's function can be written as follows
\begin{eqnarray}
\label{Gab}
G_{ab}^{-1}(k) = k^2\delta_{ab} + M^2_{ab},
\end{eqnarray}
where $M^2_{ab}$ is a constant mass matrix. 
The equations for $G_{cd}$, i.e., for the corresponding tadpoles
$Q_{ab}$, and the fields $\phi_c$ result from variations of
$V_{\scr{gHF}}$ over $G_{cd}$ and $\phi_c$,
respectively,
\begin{eqnarray}
\label{modH-SD}
&&\hspace*{-2mm}
M^2_{cd} = m^2\delta_{cd} 
\\
&&+ \frac{\lambda}{N} \left[\phi^2 \delta_{cd} 
+ 2 \phi_c\phi_d +3 Q_{aa} \delta_{cd} + 2(1-N) Q_{cd}\right]\!, 
\nonumber
\\
\label{modH-MF}
&&\hspace*{-2mm}
H_c = m^2\phi_c + \frac{\lambda}{N} \left[\phi^2\phi_c
+ Q_{aa}\phi_c + 2 Q_{cd} \phi_d\right]\!.
\end{eqnarray}
These are equations in a general nondiagonal representation. 
Applying projectors
\begin{eqnarray}
\label{pi-proj}
\Pi^\pi_{cd}&=&\frac{1}{N-1}\left(\delta_{cd} - \phi_c\phi_d/\phi^2\right),
\\
\label{sig-proj}
\Pi^\sigma_{cd}&=&\phi_c\phi_d/\phi^2, 
\end{eqnarray}
to Eq.~(\ref{modH-SD}), we project it on $\pi$ and $\sigma$ states. In
order to project the mean-field equation (\ref{modH-MF}) on the
$\sigma$-direction, we just multiply it by $\phi_c$.

In the diagonal representation ($\phi_\sigma\ne 0$, $H_\sigma = H$ and
$H_\pi=\phi_\pi=0$) these equations take the following form 
\begin{eqnarray}
\label{Phi-SD-s}
M^2_\sigma &\!\!=\!& m^2 + \frac{\lambda}{N} \left[3 \phi^2
+ (5-2N) Q_\sigma + 3(N-1) Q_\pi\right] 
\cr &\!\!=\!&
M^2_\pi + \frac{\lambda}{N} \left[2 \phi^2
+ 2(N-1) (Q_\pi - Q_\sigma)\right],
\\
\label{Phi-SD-p}
M^2_\pi &\!\!=\!& m^2 + \frac{\lambda}{N} \left[\phi^2
+ 3 Q_\sigma + (N-1) Q_\pi\right],  
\\
\label{MF-eq}
H &\!\!=\!&\phi \left[ m^2 + \frac{\lambda}{N} \left(\phi^2
+ 3 Q_\sigma + (N-1) Q_\pi\right)\right],
\end{eqnarray}
where $M^2_\pi = \Pi^\pi_{dc}M^2_{cd}$ and $M^2_\sigma =
\Pi^\sigma_{dc}M^2_{cd}$.  Here we used $Q_\sigma=Q_{\sigma\sigma}$ and
$Q_\pi=Q_{\pi\pi}$ in terms of definition (\ref{Pab}).  From these
equations it is evident that the NG theorem is fulfilled. Indeed, in the
phase of spontaneously broken symmetry ($H=0$) the square-bracketed term of
the field equation (\ref{MF-eq}) equals zero, which is precisely the pion
mass, cf. Eq. (\ref{Phi-SD-p}). At the same time, as it has been
demonstrated in numerous papers (see, e.g.,
Refs.~\cite{Baym-Grin,HM65,Aouissat,Lenaghan}), the solution of the
conventional HF set of equations (\ref{Phi-J})--(\ref{Phi-S}), i.e.,
without $\Delta\Phi$, violates the NG theorem. A detailed analysis of the
conventional HF equations in a notation similar to ours has been given in
Ref.~\cite{Lenaghan}.

\section{Renormalization of the ${\mbox{\bf gHF}}$ Approximation}

Significant progress in proper renormalization of $\Phi$-derivable
approximations was recently achieved in Refs.~\cite{HK3,HK1,Reinosa1}.
Here we follow the renormalization scheme of Ref.~\cite{HK3}, i.e.,
that constructed precisely for the conventional HF approximation to
$\lambda \phi^4$ theory in the $O(N)$ broken phase.  This
renormalization is based on the BPHZ formalism.

\subsection{Equations of Motion}
\label{Equations of Motion}

The equations of motion (\ref{Phi-SD-s})--(\ref{MF-eq}) involve 
tadpole terms which, based on the explicit form of the Green's function
(\ref{Gab}), can be written as 
\begin{eqnarray}
\label{Q(n)}
Q_a = \int \frac{d^3 k}{(2\pi)^3}\frac{1}{\epsilon_a({\vec k})}
\left[n\left(\epsilon_a({\vec k})\right)+\frac{1}{2}\right],
\end{eqnarray}
where $\epsilon_a({\vec k})=({\vec k}^2+M_a^2)^{1/2}$ and 
\begin{eqnarray}
\label{n(p)}
n(\epsilon) = \frac{1}{\exp(\epsilon/T)-1} 
\end{eqnarray}
is the thermal occupation number. Evidently, the $Q$-function
consists of two parts
\begin{eqnarray}
\label{Q-nonren}
Q_a = Q_a^T + Q_a^{\scr{(div)}},
\end{eqnarray}
where
\begin{eqnarray}
\label{P}
Q_a^T = \int \frac{d^3 k}{(2\pi)^3} \frac{1}{\epsilon_a({\vec k})} 
n\left(\epsilon_a({\vec k})\right) 
\end{eqnarray}
is the convergent thermal part of the tadpole, which is finite, and the
divergent part 
\begin{eqnarray}
\label{P-div}
Q_a^{\scr{(div)}} &=& \frac{1}{2} 
\int \frac{d^3 k}{(2\pi)^3} \frac{1}{\epsilon_a({\vec k})} 
\cr
&=&
\frac{M_a^2}{(4\pi)^2}\left( -\frac{1}{\epsilon} +
  \ln\frac{M_a^2}{\mu^2} - 1 \right)
\end{eqnarray}
is regularized within dimensional regularization. Here $\epsilon\to 0$, and
$\mu$ is a regularization scale. We apply the same mass independent
renormalization conditions as in Ref. \cite{HK3}, i.e., that in the
\emph{symmetric vacuum} the self-energies vanish~\cite{Kugo}
\begin{eqnarray}
\label{renorm-cond}
\Sigma_a (T=0,\phi=0,m^2=\mu^2>0) = 0, \\
\partial_{m^2} \Sigma_a (T=0,\phi=0,m^2=\mu^2>0)=0,
\end{eqnarray}
for all $a$. Such a renormalization scheme preserves the O($N$) symmetry of
the model~\cite{Kugo}. Since $\Sigma_a$ is momentum independent in the
approximation under consideration, additional momentum-derivative
conditions are not required. Upon application of this scheme, Eqs.
(\ref{Phi-SD-s})--(\ref{MF-eq}) keep their form with the $Q_a$ quantities
substituted by the renormalized tadpoles.
\begin{eqnarray}
\label{Q-ren}
\hspace*{-7mm}
Q_a^{\scr{(ren)}} &=& Q_a^T 
\cr
\hspace*{-7mm}
&+& 
\frac{1}{(4\pi)^2}\left[
M_a^2\left(\ln\frac{M_a^2}{\mu^2} - 1\right) +\mu^2\right].
\end{eqnarray}
As it was shown in Ref.~\cite{HK3}, this renormalization description
requires only vacuum (temperature independent) counter terms. For the sake
of further discussion, note that according to Ref.~\cite{HK3} the
renormalization of the conventional HF approximation results in precisely
the same equations as those in the CT scheme of Ref.~\cite{Lenaghan}, in
spite of the different approaches used. The renormalized conventional HF
approximation was thoroughly studied in~\cite{Lenaghan}. Therefore, those
results are very useful for comparison with the present treatment.

\subsection{Effective Potential}
\label{Potential}

The thermodynamic potential (\ref{V-H}) is renormalized following the
procedure outlined in Ref.~\cite{HK1}. In the gHF approximation
complications arise only in the $\ln\det G^{-1}(k)$ term.  Because of the
topology of the diagrams used in the gHF approximation, for all other
contributions to the effective action, we only have to insert the already
renormalized self-energies, i.e., $Q_a^{\scr{(ren)}}$ tadpoles of Eq.
(\ref{Q-ren}), to renormalize them.  This is legitimate, because we do not
encounter any additional contributions from subdivergences, cf.
Ref.~\cite{HK1}. In the diagonal representation the remaining part to be
renormalized takes the form
\begin{eqnarray}
\label{ln-term}
\frac{1}{2} \int_\beta d^4 k \ln\det G^{-1}(k)=
L_\sigma + (N-1) L_\pi, 
\end{eqnarray}
where we have introduced the brief notation
\begin{eqnarray}
\label{L-def}
\hspace*{-7mm}
L_a&=&\frac{1}{2} \int_\beta d^4 k \ln G_a^{-1}(k)
\cr
\hspace*{-7mm}
&=&
\int\frac{d^3 k}{(2\pi)^3}
\left\{ \frac{\epsilon_a}{2} + 
T\ln\left[1-\exp\left(-\frac{\epsilon_a}{T}
\right)\right]\right\}. 
\end{eqnarray}
The $L_a$ also consists of two parts: the convergent thermal
part 
\begin{eqnarray}
\label{L-T}
L_a^T=T
\int\frac{d^3 k}{(2\pi)^3}
\ln\left[1-\exp\left(-\frac{\epsilon_a({\vec k})}{T}
\right)\right], 
\end{eqnarray}
which is finite, and the divergent integral
\begin{eqnarray}
\label{L-div}
L_a^{\scr{(div)}}=
\int\frac{d^3 k}{(2\pi)^3}
\frac{\epsilon_a({\vec k})}{2}. 
\end{eqnarray}
This expression implicitly depends on temperature through the mass $M_a$.
We regularize it by means of a momentum cut-off $\Lambda$:
\begin{widetext}
\begin{eqnarray}
\label{L-reg}
\hspace*{-8mm}
L_a^{\scr{(reg)}}(M_a)&=&
\int\frac{d^3 k}{(2\pi)^3}
\frac{\epsilon_a({\vec k})}{2} 
\Theta\left(\Lambda^2-\epsilon_a^2({\vec k})\right)
=
\frac{1}{3(2\pi)^2}
\left(
\left(\Lambda^2-M_a^2\right)^{3/2}\Lambda \vphantom{\frac{1}{8}}\right.
\cr
\hspace*{-8mm}
&-&
\left.
\frac{1}{8}
\left[2\Lambda^3\sqrt{\Lambda^2-M^2}-5\Lambda M^2\sqrt{\Lambda^2-M^2}-
3M^4\ln M+3M^4\ln\left(\Lambda+\sqrt{\Lambda^2-M^2}\right)\right]
\right). 
\end{eqnarray}
To renormalize the effective potential, we use a mass-independent
renormalization scheme in order to avoid effects of unphysical IR
singularities. We impose the following renormalization conditions on the
effective potential
\begin{eqnarray}
\label{V-ren-c1}
\hspace*{-8mm}
V^{\scr{(ren)}}\left(T\!=\!0,\phi\!=\!0,m^2\right)
\left|_{m^2=\mu^2>0}\right.&=&0,
\\
\label{V-ren-c2}
\hspace*{-7mm}
\partial_{m^2}V^{\scr{(ren)}}\left(T\!=\!0,\phi\!=\!0,m^2\right)
\left|_{m^2=\mu^2>0}\right.&=&0,
\\
\label{V-ren-c3}
\hspace*{-7mm}
\partial^{2}_{m^2}V^{\scr{(ren)}}\left(T\!=\!0,\phi\!=\!0,m^2\right)
\left|_{m^2=\mu^2>0}\right.&=&0.
\end{eqnarray}
We need precisely these three conditions to remove all the divergences.
Imposing these conditions, we keep in mind that other parts of the
effective potential, except for $L_a$, have already been renormalized such
that they fulfill these conditions on their own. This leads to
\begin{eqnarray}
\label{L-ren}
L_a^{\scr{(ren)}}&=&
L_a^T+\underset{\Lambda\rightarrow \infty}{\lim}
\left[
  L_a^{\scr{(reg)}}(M_a)-L_a^{\scr{(reg)}}(\mu)
-(M_a^2-\mu^2)
\frac{\partial L_a^{\scr{(reg)}}(\mu)}{\partial \mu^2}-
\frac{1}{2}(M_a^2-\mu^2)^2
\frac{\partial^2 L_a^{\scr{(reg)}}(\mu)}{\partial (\mu^2)^2}
\right] 
\cr  
&=&
L_a^T- 
\frac{1}{128\pi^{2}}\left(
  3M^{4}-4M^2\mu^{2}+\mu^{4}-2M^{4}\ln\frac{M^2}{\mu^2}
 \right).  
\end{eqnarray}
In terms of these $L_a^{\scr{(ren)}}$ and $Q_a^{\scr{(ren)}}$ the
renormalized  effective potential reads
\begin{eqnarray}
\label{V-ren}
&&V_{\scr{gHF}}^{\scr{(ren)}}(\phi,T)=\frac{1}{2}m^2\phi^2
+\frac{\lambda}{4N}\phi^4- H\phi
+L_\sigma^{\scr{(ren)}}+(N-1)L_\pi^{\scr{(ren)}}
\cr
&&+
\frac{1}{2}\left[
m^2\left(Q_\sigma^{\scr{(ren)}}+(N-1)Q_\pi^{\scr{(ren)}}\right)
-
M_\sigma^2Q_\sigma^{\scr{(ren)}}-(N-1)M_\pi^2Q_\pi^{\scr{(ren)}}
+
\frac{\lambda}{N}\phi^2
\left(3Q_\sigma^{\scr{(ren)}}+(N-1)Q_\pi^{\scr{(ren)}}\right)
\right]
\cr
&&+
\frac{\lambda}{4N}\left[
3\left(Q_\sigma^{\scr{(ren)}}
+(N-1)Q_\pi^{\scr{(ren)}}\right)^2
-2(N-1)\left(
\left[Q_\sigma^{\scr{(ren)}}\right]^2
+(N-1)\left[Q_\pi^{\scr{(ren)}}\right]^2
\right)
\right].
\end{eqnarray}

\subsection{Vacuum ($T = 0$)}
\label{Vacuum}

At $T=0$, the quantities under investigation are ``experimentally''
known\footnote{These values are relevant for the case $N=4$.}:
$M_\pi(T=0)=m_\pi=139$~MeV, $M_\sigma (T=0)=m_\sigma=600$~MeV, and the pion
decay constant $\phi_0=f_\pi=93$~MeV. At $T = 0$ these known quantities
should satisfy Eqs.~(\ref{Phi-SD-s})--(\ref{MF-eq}) with renormalized
tadpoles $Q_a^{\scr{(ren)}}$, cf. Eq. (\ref{Q-ren}),
\begin{eqnarray}
\label{Phi-SD-s-0}
\hspace*{-8mm}
m^2_\sigma &\!=\!& m^2_\pi 
+
\frac{2\lambda}{N} f_\pi^2
+
\frac{2(N-1)\lambda}{(4\pi)^2N} \left[
m_\pi^2\left(\ln\frac{m_\pi^2}{\mu^2}-1\right)
-
m_\sigma^2\left(\ln\frac{m_\sigma^2}{\mu^2}-1\right)
\right]\!,
\\
\label{Phi-SD-p-0}
\hspace*{-8mm}
m^2_\pi &\!=\!& m^2 
+
\frac{\lambda}{N} f_\pi^2
+ 
\frac{\lambda}{(4\pi)^2N} \left[
(N+2)\mu^2
+
3 m_\sigma^2\left(\ln\frac{m_\sigma^2}{\mu^2}-1\right)
+(N-1) m_\pi^2\left(\ln\frac{m_\pi^2}{\mu^2}-1\right)
\right]\!,
\\
\label{MF-eq-0}
\hspace*{-8mm}
\frac{H}{f_\pi}&\!=\!& m^2 
+
\frac{\lambda}{N} f_\pi^2
+ 
\frac{\lambda}{(4\pi)^2N} \left[
(N+2)\mu^2
+
3 m_\sigma^2\left(\ln\frac{m_\sigma^2}{\mu^2}-1\right)
+(N-1) m_\pi^2\left(\ln\frac{m_\pi^2}{\mu^2}-1\right)
\right]\!.  
\end{eqnarray}
In order to make this set of equations consistent, we should put
\begin{eqnarray}
\label{H-pi}
H=m^2_\pi f_\pi, 
\end{eqnarray}
i.e., precisely the same as at the tree level. Then Eqs.~(\ref{Phi-SD-p-0})
and (\ref{MF-eq-0}) become identical. Now we can solve for the remaining
equations, (\ref{Phi-SD-s-0}) and (\ref{Phi-SD-p-0}), and express the
renormalized quantities $m^2$ and $\lambda$ in terms of physical quantities
$m_\pi$, $m_\sigma$, and $f_\pi$
\begin{eqnarray}
\label{lambda}
\hspace*{-5mm}
\lambda &=&
N\left(m^2_\sigma - m^2_\pi \right)
\left\{
2f_\pi^2
+ 
\frac{2(N-1)}{(4\pi)^2} \left[
m_\pi^2\left(\ln\frac{m_\pi^2}{\mu^2}-1\right)
-m_\sigma^2\left(\ln\frac{m_\sigma^2}{\mu^2}-1\right)
\right]
\right\}^{-1},  
\\
\label{mass}
\hspace*{-5mm}
-m^2 &=& -m^2_\pi 
+
\frac{\lambda}{N} f_\pi^2
+ 
\frac{\lambda}{(4\pi)^2N} \left[
(N+2)\mu^2
+3 m^2_\sigma\left(
\ln\frac{m_\sigma^2}{\mu^2}-1
\right)
+(N-1) m^2_\pi\left(
\ln\frac{m_\pi^2}{\mu^2}-1
\right)
\right].
\end{eqnarray}
\end{widetext}
The expressions manifestly show that the performed renormalization is
$\mu$-scale independent in the vacuum. Indeed, $m^2$ and $\lambda$ can take
any values depending on the choice of the renormalization scale $\mu$,
while the same observables keep their physical values at any
renormalization scale $\mu$ (within a certain range of $\mu$).  In this
respect, the gHF approximation is similar to the leading-order
$1/N$-approximation, where also both the NG theorem and scale-independent
renormalization in the vacuum are fulfilled~\cite{Lenaghan}.  The
restriction to a certain range is related to the conditions $\lambda>0$ and
$m^2<0$ which should be met. At the scale $\mu_0$, determined by the
equation
\begin{eqnarray}
\label{mu-0}
f_\pi^2
&+ \; 
\frac{(N-1)}{(4\pi)^2}&
\left(
m_\pi^2\ln\frac{m_\pi^2}{\mu_0^2}-m_\pi^2
\right.
\cr
&&
\hspace*{3mm}
\left.
-m_\sigma^2\ln\frac{m_\sigma^2}{\mu_0^2}+m_\sigma^2
\right)
=0,
\end{eqnarray}
$\lambda$ and $m^2$ become singular, cf. Eqs. (\ref{lambda}) and
(\ref{mass}). Moreover, $\mu<\mu_0$ implies $m^2>0$ and $\lambda<0$,
which defers a spontaneously broken phase and makes the
theory unstable because of uncompensated attraction
($\lambda\phi^4<0$). Therefore, the range of scale independent
renormalization in the vacuum is restricted from below by
\begin{eqnarray}
\label{mu-0rest}
\mu>\mu_0. 
\end{eqnarray}
The situation is completely different within the conventional HF
approximation. In that case, physical observables do depend on the
renormalization scale and can be reproduced only at a single value of the
scale $\mu^2=m_\sigma^2/e$, as demonstrated in~\cite{Lenaghan}. The reason
of this difference is that the conventional HF approximation
of~\cite{Lenaghan} violates the NG theorem at $H=0$.  This stems from the
fact that the equations for the pion self-energy and the mean field are not
identical in the broken phase and thus leave three nondegenerate equations
for three quantities $m^2$, $\lambda$ and $\mu$, from which the $\mu$ value
is unambiguously determined. In our case, this set of equations is
degenerate, i.e., the equations for the pion self-energy and the mean field
are identical as a consequence of the fulfilled NG theorem. This gives us
freedom for an arbitrary choice of $\mu$. Thus, we arrive at an important
conclusion which concerns all partial resummation schemes applied to the
case of spontaneously broken symmetry: a scale-independent renormalization
in the vacuum is possible only if the scheme preserves the NG theorem.
This is an important aspect with respect to possible
renormalization-group considerations for this type of self consistent
approximations.

However, the scale dependence still persists at finite temperature, which
is already seen from the analysis of the symmetry-restoration points. 

\subsection{Symmetry Restoration Points at $H=0$}

Starting from the broken phase, where $M^2_\pi=0$ and
$\phi^2>0$, there exists a temperature range $T_R\in[T_1,T_2]$,
cf. Fig. \ref{fig-zm:mp0-600} below,
where the classical field vanishes together with the pion mass
\begin{eqnarray}
\label{R_point}
M^2_\pi(T_R) = \phi^2(T_R) = 0.   
\end{eqnarray}
This precisely occurs, when the two equations (\ref{Phi-SD-p}) and
(\ref{MF-eq}) reduce to a single one. Solving for this single equation
with renormalized $Q_a^{\scr{(ren)}}$ tadpoles
(\ref{Q-ren}) gives
\begin{eqnarray}
\label{TR}
\hspace*{-7mm} 
&&T_R^2 = \frac{12}{(N+2)} 
\left[
\left(1+\frac{3}{(N-1)}\frac{M_\sigma^2(T_R)}{m_\sigma^2}\right)
f_\pi^2
\right.  
\cr
\hspace*{-7mm} 
&&+ \!
\left.
\frac{3m_\sigma^2}{(4\pi)^2}\!
\left(\ln\frac{m_\sigma^2}{\mu^2} -1\right) 
\left(1-\frac{M_\sigma^2(T_R)}{m_\sigma^2}\right)
\right],   
\end{eqnarray}
where we have used Eqs. (\ref{lambda}) and (\ref{mass}) to express $m^2$
and $\lambda$ in terms of the vacuum mass $m_\sigma$.  Strictly speaking,
this is not a solution for $T_R$, since the r.h.s. of (\ref{TR}) still
depends on $T_R$ through $M_\sigma^2(T_R)$. Nevertheless, this expression
is already quite simple to analyze.

The lower bound $T_1$ of condition (\ref{R_point}) is of central importance
for the phase transition in the conventional HF-approximation \cite{IRK05}
but of minor relevance in the gHF-scheme, since it corresponds to the
metastable solution, cf. Sect.~\ref{Results} below. It is determined if
simultaneously also $M^2_\sigma=0$ occurs
\begin{eqnarray}
\label{Tc}
\hspace*{-9mm} 
T_1^2 &=& \frac{12}{(N+2)} 
\left[f_\pi^2
+ \frac{3m_\sigma^2}{(4\pi)^2}\left( 
\ln\frac{m_\sigma^2}{\mu^2} -1\right) 
\right]\!.  
\end{eqnarray}
It is still $\mu$-dependent, in spite of the scale-independent
renormalization in the vacuum. At large $\mu$, i.e., above some $\mu_1$,
$T_1^2$ can even become negative, which means that then this solution does
not exist.

For the stable solution of the gHF approximation it numerically occurs that
$M_\sigma(T_R)\approx m_\sigma$, cf. Ref.~\cite{IRK05} and
Sect.~\ref{Results} below. This almost removes the $M_\sigma(T_R)$
dependence from the r.h.s. of Eq. (\ref{TR}) and makes the corresponding
temperature $T_2$ almost $\mu$-independent
\begin{eqnarray}
\label{T2}
\hspace*{-7mm} 
T_2^2 \simeq \frac{12}{(N-1)} f_\pi^2. 
\end{eqnarray}
This value coincides with that of the naive renormalization~\cite{IRK05}.
The solution $T_2$ corresponds to a partial symmetry restoration, since
here we still have $M_\sigma(T_2)\neq M_\pi(T_2)$ in spite of $\phi=0$.

\section{Results for $\mathbf{ N= 4}$}
\label{Results}

For the numerical calculations we use the following parameters:
$m_\sigma=600$~MeV and $f_\pi=93$~MeV. The pion mass is either zero,
$m_\pi=0$, in the case of exact symmetry, or $m_\pi=139$~MeV for the
approximate symmetry.  The general structure of the solutions to the
renormalized Eqs.~(\ref{Phi-SD-s})--(\ref{MF-eq}) is similar to that
obtained with the naive renormalization \cite{IRK05} but with extra
complications caused by the additional dependence on the renormalization
scale $\mu$. According to Eq. (\ref{mu-0}) physically reasonable solutions
exits only for $\mu>\mu_0$, where $\mu_0$ as the solution of Eq.
(\ref{mu-0}) equals $\simeq 200$~MeV for the above specified parameters.

There are several different branches of the solution. Stable and physically
meaningful are determined by the principle of maximum pressure, the
pressure being given by the effective potential (\ref{V-ren})
\begin{eqnarray}
\label{pressure}
P = -V_{\scr{gHF}}^{\scr{(ren)}}(\phi,T)+\mathrm{const.}   
\end{eqnarray}
Here the constant is determined by the condition that the pressure
should vanish for the physical vacuum, i.e., in the spontaneously
broken phase, while our renormalization condition (\ref{V-ren-c1})
determines $V_{\scr{gHF}}^{\scr{(ren)}}$ to be zero in the unphysical,
symmetric vacuum.

\subsection{Exact $O(4)$ Symmetry}
\label{Exact}

\begin{figure}[ht]
\includegraphics[width=5.5cm,angle=-90]{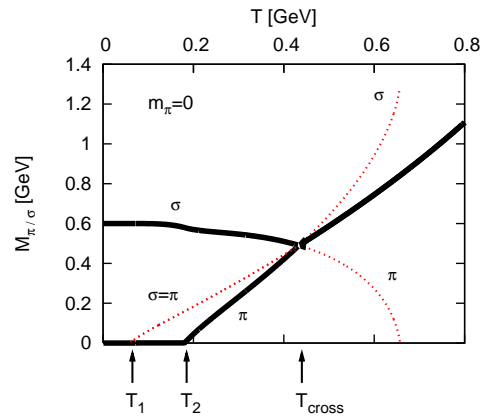}
\caption{Meson masses as functions of temperature for $\mu=$ 600 MeV and
  $m_\pi=0$ case. Stable branch is presented by solid lines, whereas the
  metastable one -- by dashed lines.}
\label{fig-zm:mp0-600}
\end{figure}

\begin{figure}[ht]
\includegraphics[width=5.5cm,angle=-90]{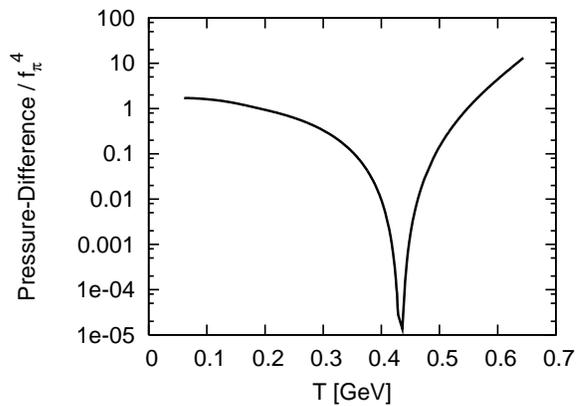}
\caption{The same as in Fig. \ref{fig-zm:mp0-600} but for the pressure
  difference between stable and metastable branches.}
\label{fig-dp:mp0-600}
\end{figure}

\begin{figure}[ht]
\includegraphics[width=5.5cm,angle=-90]{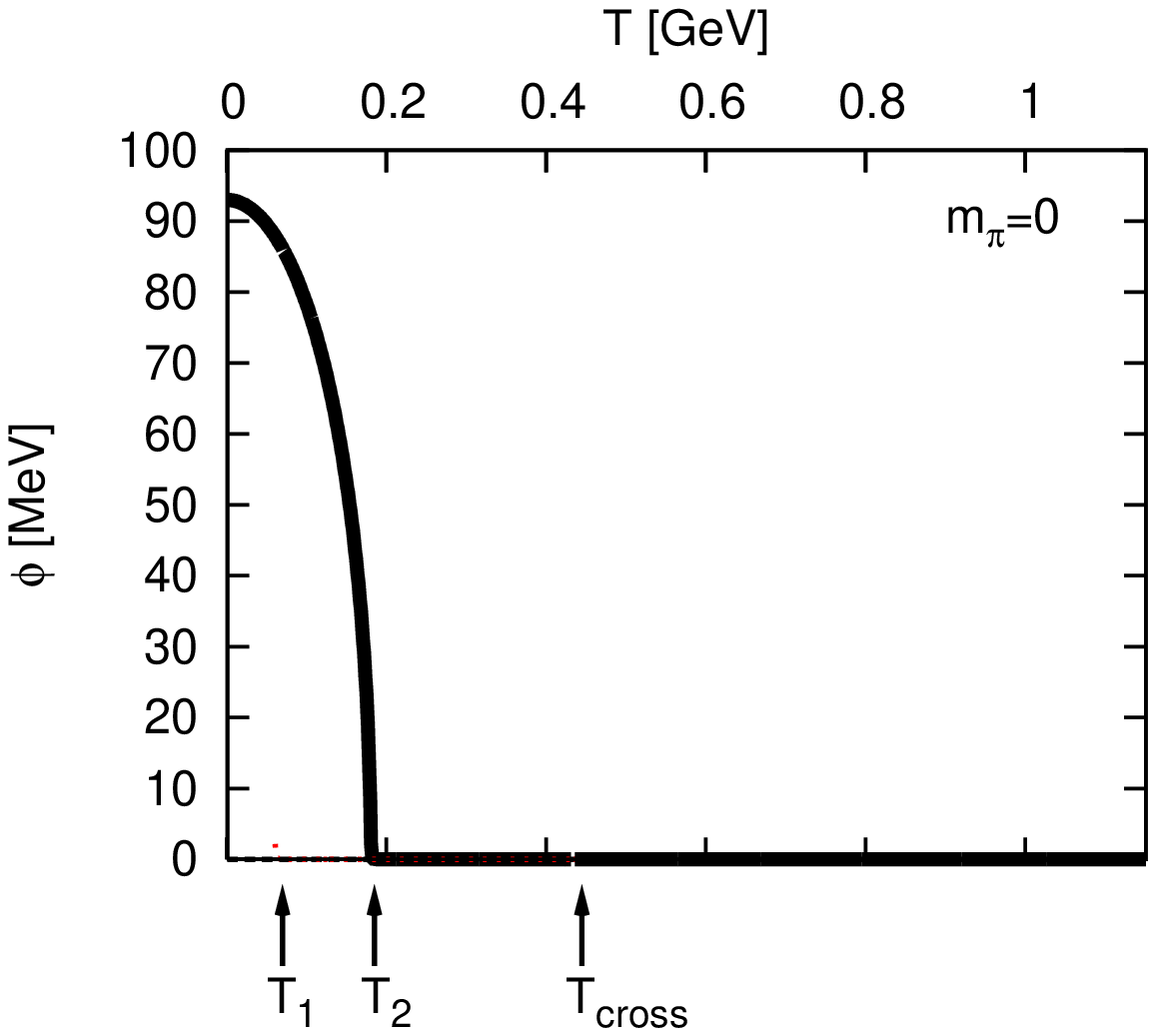}
\caption{The same as in Fig. \ref{fig-zm:mp0-600} but for the field $\phi$.
}
\label{fig-f:mp0-600}
\end{figure}

\begin{figure}[ht]
\includegraphics[width=5.5cm,angle=-90]{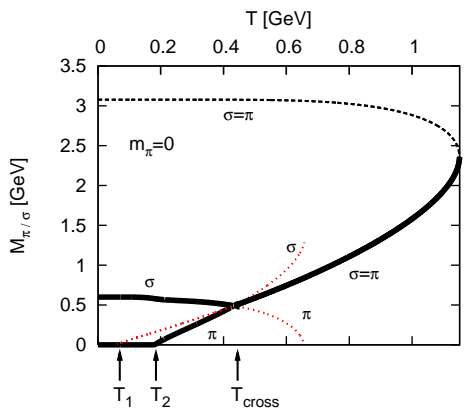}
\caption{Meson masses of Fig. \ref{fig-zm:mp0-600} but in wider temperature
  region. The upper metastable branch is displayed by the long-dashed
  line.}
\label{fig-m:mp0-600}
\end{figure}

\begin{figure}[ht]
\includegraphics[width=5.5cm,angle=-90]{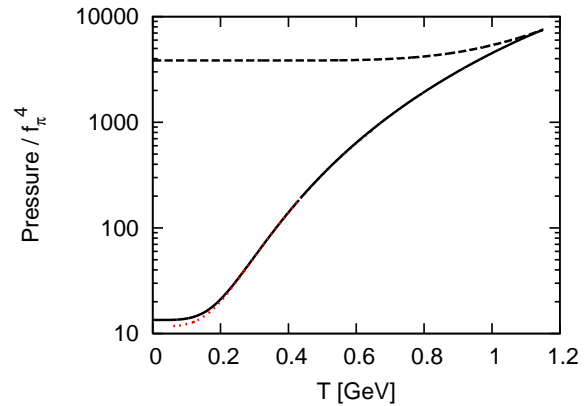}
\caption{The same as in Fig. \ref{fig-m:mp0-600} but for the pressure.} 
\label{fig-p:mp0-600}
\end{figure}

The actual structure of the solution depends on the renormalization scale
$\mu$. We start with a moderate scale $\mu=600$~MeV, i.e., of the order of
$m_\sigma$. The results are presented in
Figs.~\ref{fig-zm:mp0-600}--\ref{fig-p:mp0-600}.

In the narrow temperature range, displayed in Fig. \ref{fig-zm:mp0-600},
the results are qualitatively similar to those obtained with the naive
renormalization~\cite{IRK05}. The stable branch starts at $T=0$ from the
physical vacuum values for the masses and the classical field and crosses
the metastable branch at $T_{\scr{cross}}\approx 440$~MeV. In terms of the
pressure, they are touching rather than crossing (see
Fig.~\ref{fig-dp:mp0-600}). Therefore, no transition from one branch to
another occurs at $T_{\scr{cross}}$. In the broken-symmetry phase, the pion
mass equals zero.  Then a phase transition of the second order occurs at
$T_2\simeq 180$~MeV, at which the field becomes zero (see Fig.
\ref{fig-f:mp0-600}). However, the $\pi$ and $\sigma$ masses still differ
beyond this transition point.  They become equal only after a second phase
transition, which is also of second order, at $T_{\scr{cross}}$. Note that
the equal-mass solution above $T_{\scr{cross}}$ is precisely the same as in
the conventional HF approximation (cf. Refs. \cite{HK3,Lenaghan}), since
the gapless modification term (\ref{dPhi-ab}) vanishes in this case. The
$T_1$ point proves to be irrelevant for the stable branch. Rather it is the
starting point for the metastable branch, which in the range of
$T_1<T<T_{\scr{cross}}$ precisely coincides with the solution of the
conventional HF approximation \cite{HK3,Lenaghan}.  The corresponding field
is always zero for this branch. Contrary to the case of the naive
renormalization \cite{IRK05}, this metastable branch ends at some
temperature ($\approx 650$~MeV).

\begin{figure}[ht]
\includegraphics[width=5.5cm,angle=-90]{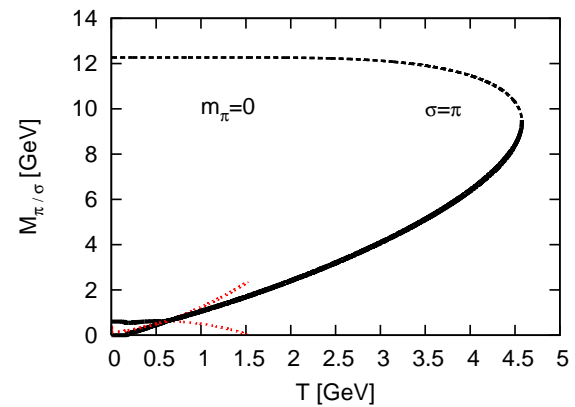}
\caption{Meson masses as functions of temperature for $\mu=1200$~MeV and
  $m_\pi=0$ case. Stable branches are presented by solid lines, whereas
  metastable ones by dashed lines.}
\label{fig-m:mp0-1200}
\end{figure}
\begin{figure}[ht]
\includegraphics[width=5.5cm,angle=-90]{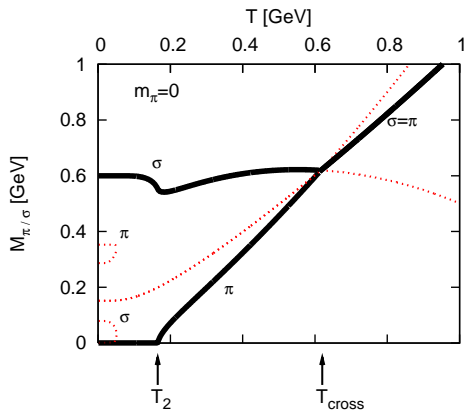}
\caption{Zoomed low-temperature region of Fig. \ref{fig-m:mp0-1200}.}
\label{fig-zm:mp0-1200}
\end{figure}
\begin{figure}[ht]
\includegraphics[width=5.5cm,angle=-90]{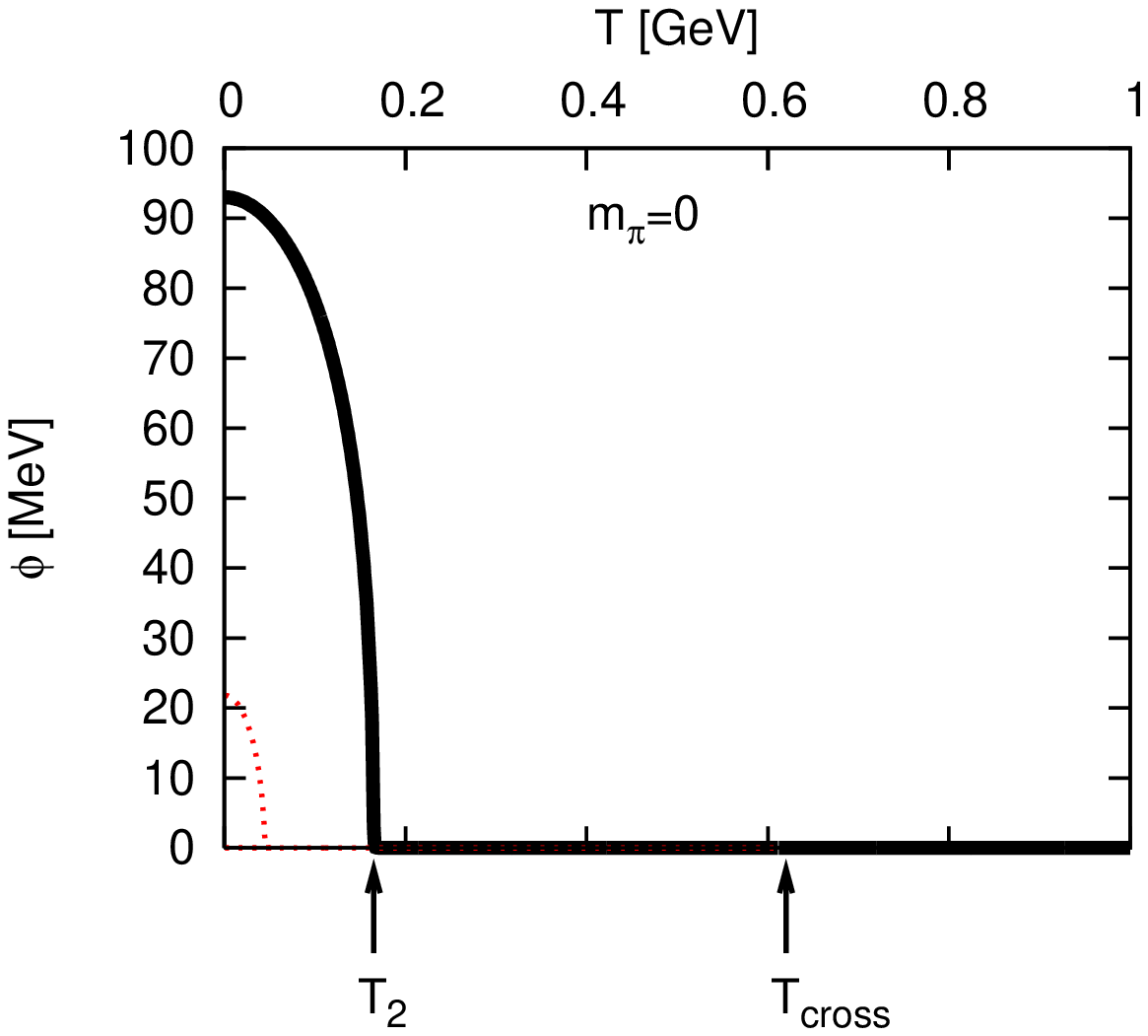}
\caption{The same as in Fig. \ref{fig-m:mp0-1200} but for the field
  $\phi$. }  
\label{fig-f:mp0-1200}
\end{figure}

However, in the wider temperature range, displayed in Fig.
\ref{fig-m:mp0-600}, we see a significant difference from the results
obtained with the naive renormalization \cite{IRK05}.  Neither stable nor
metastable solutions exist above a certain temperature $T_{\scr{end}}$
which is $\approx 1.15$~GeV for the considered $\mu=600$~MeV.  Let us
remind that this solution above $T_{\scr{cross}}$ corresponds to the
conventional HF approximation.  Therefore, it is precisely the same as in
previous conventional HF calculations with
renormalization~\cite{Lenaghan,HK3}. The occurrence of such an end point in
the HF approximation was first pointed out by Baym and
Grinstein~\cite{Baym-Grin}.

At $T_{\mathrm{end}}$, the stable branch of the solution joins the upper
branch. This upper branch at any temperature corresponds to equal masses
and zero field, i.e., $M_\pi=M_\sigma$ and $\phi=0$, and hence is also a
solution to the conventional renormalized HF approximation. It starts with
very high values of masses ($\approx 3$~GeV) in the vacuum and also ends at
$T_{\scr{end}}$. The vacuum pressure for this branch, evaluated according
to Eq.~(\ref{pressure}), is rather high (see Fig.  \ref{fig-p:mp0-600}).

It is important note that for the upper branch and at temperatures near the
endpoint the logarithmic terms in the gap equation, $\propto
\ln(M_{\pi/\sigma}^2/\mu^2)$, become large. This indicates that at such
points the expansion of the $\Phi$ functional in powers of the renormalized
coupling becomes unreliable, because the effective coupling becomes large.
Therefore we consider the upper branch not a physically meaningful
solution. The same holds true for temperatures close to the endpoint
temperature. Such a behavior must be expected for any effective theory and
was indeed also observed in Quantum Hadro Dynamics (QHD) in~\cite{sewal97}.

At larger renormalization scales, $\mu$, the global pattern of the solution
remains qualitatively similar, as seen from Fig. \ref{fig-m:mp0-1200}. Only
$T_{\scr{cross}}$ and $T_{\scr{end}}$ move to higher temperatures.
Inspecting the low-temperature region in more detail, cf.  Fig.
\ref{fig-zm:mp0-1200}, we see that a new metastable solution, which ends
already at rather low temperature, appears. This new solution has a nonzero
field (Fig. \ref{fig-f:mp0-1200}) but violates the NG theorem. In addition
the metastable solution, which before started at $T_1$, now begins at zero
temperature, since at $\mu=$ 1.2 GeV we have already $T_1^2<0$, cf. Eq.
(\ref{Tc}).

\begin{figure}[ht]
\includegraphics[width=5.5cm,angle=-90]{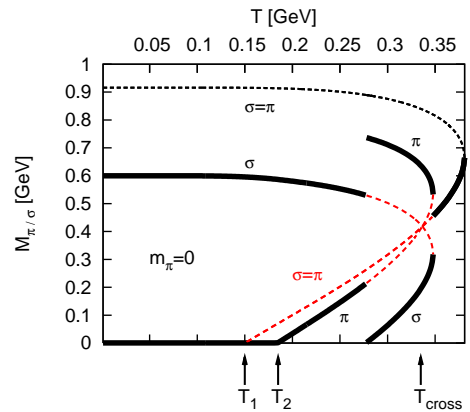}
\caption{Meson masses  as
  functions of temperature for $\mu=$ 300 MeV and $m_\pi=0$
  case. Stable branches are 
  presented by solid lines, whereas metastable ones -- by 
  dashed lines.} 
\label{fig-m:mp0-300}
\end{figure}
\begin{figure}[ht]
\includegraphics[width=5.5cm,angle=-90]{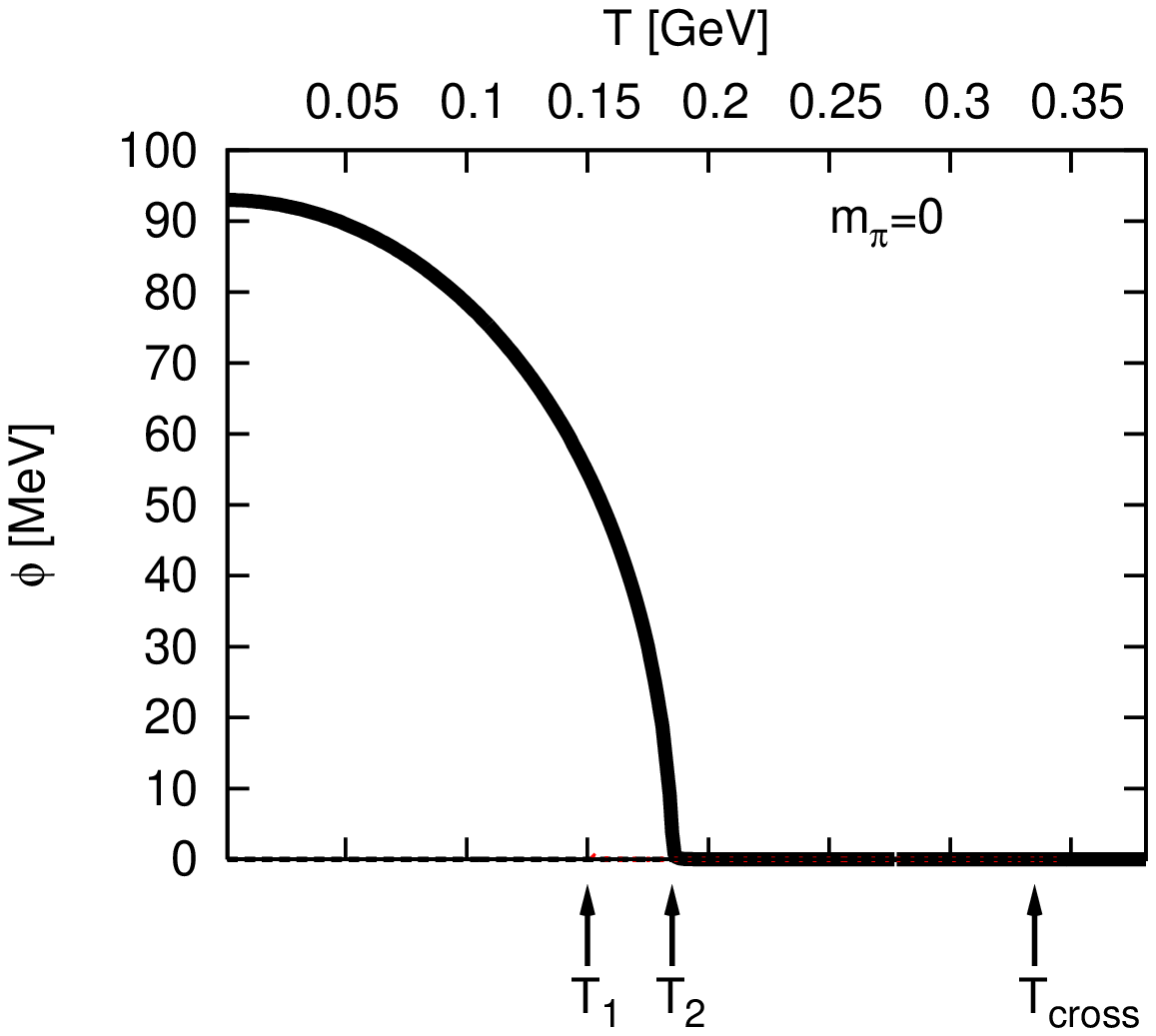}
\caption{The same as in Fig. \ref{fig-m:mp0-300} but for the field $\phi$.
}
\label{fig-f:mp0-300}
\end{figure}
\begin{figure}[ht]
\includegraphics[width=5.5cm,angle=-90]{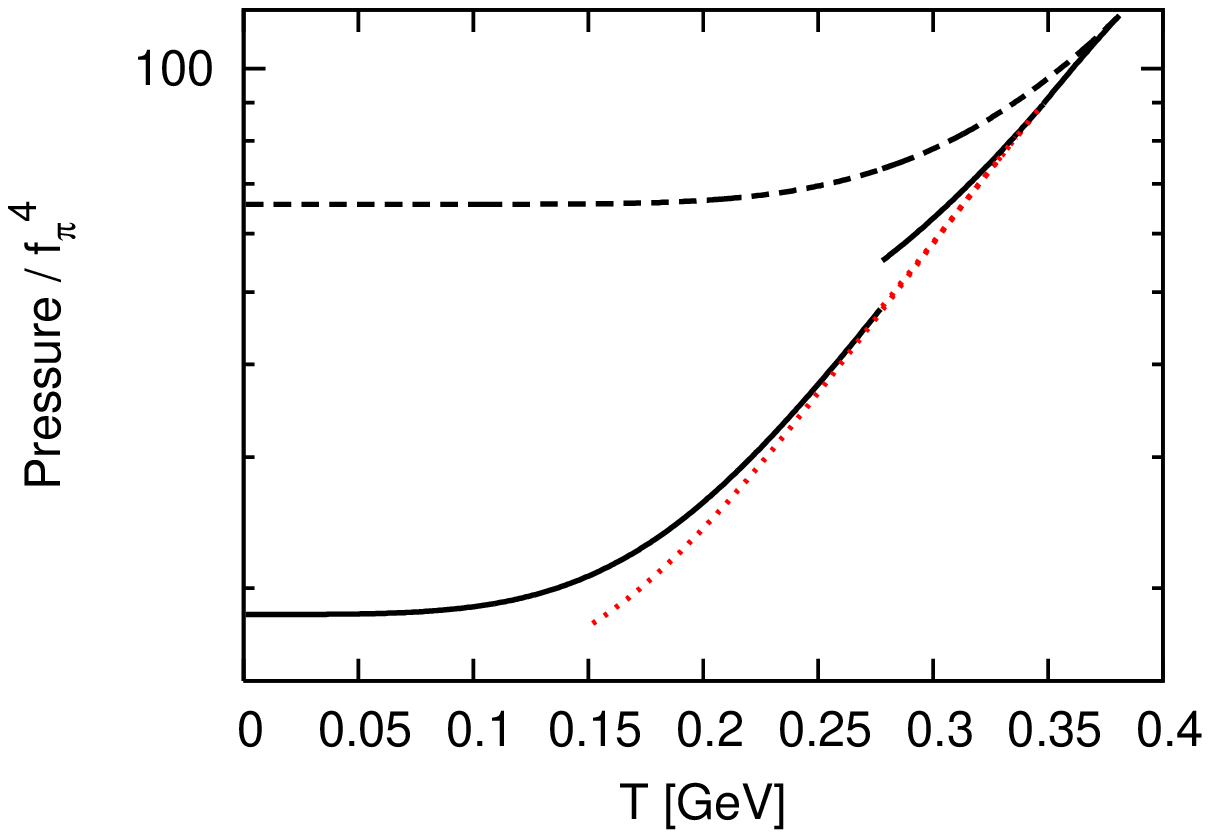}
\caption{The same as in Fig. \ref{fig-m:mp0-300} but for the pressure.}
\label{fig-p:mp0-300}
\end{figure}

If we take a low value for the scale $\mu$, the structure of the stable
solution becomes more involved, see
Figs.~\ref{fig-m:mp0-300}--\ref{fig-p:mp0-300}. In the broken-symmetry
sector we still have a massless pion and a nonzero field, see
Fig.~\ref{fig-f:mp0-300}. However, at higher temperatures the metastable
branch, displayed by the dotted line, reveals back-bending. This
back-bended part of the branch turns out to be the most stable one, see
Fig.~\ref{fig-p:mp0-300}.  As a result we arrive at a complicated structure
for the stable solution, where even the pressure turns out to be
discontinuous.

A common feature for all scales is that the point of the first phase
transition, $T_2\simeq 180$~MeV, is approximately $\mu$-independent and has
about the same value as that in the naive renormalization scheme
\cite{IRK05}. At the same time, the point of the second phase transition,
$T_{\scr{cross}}$, and the end point of the stable solution,
$T_{\scr{end}}$, are essentially $\mu$-dependent.

\subsection{Approximate $O(4)$ Symmetry}

\begin{figure}[ht]
\includegraphics[width=5.5cm,angle=-90]{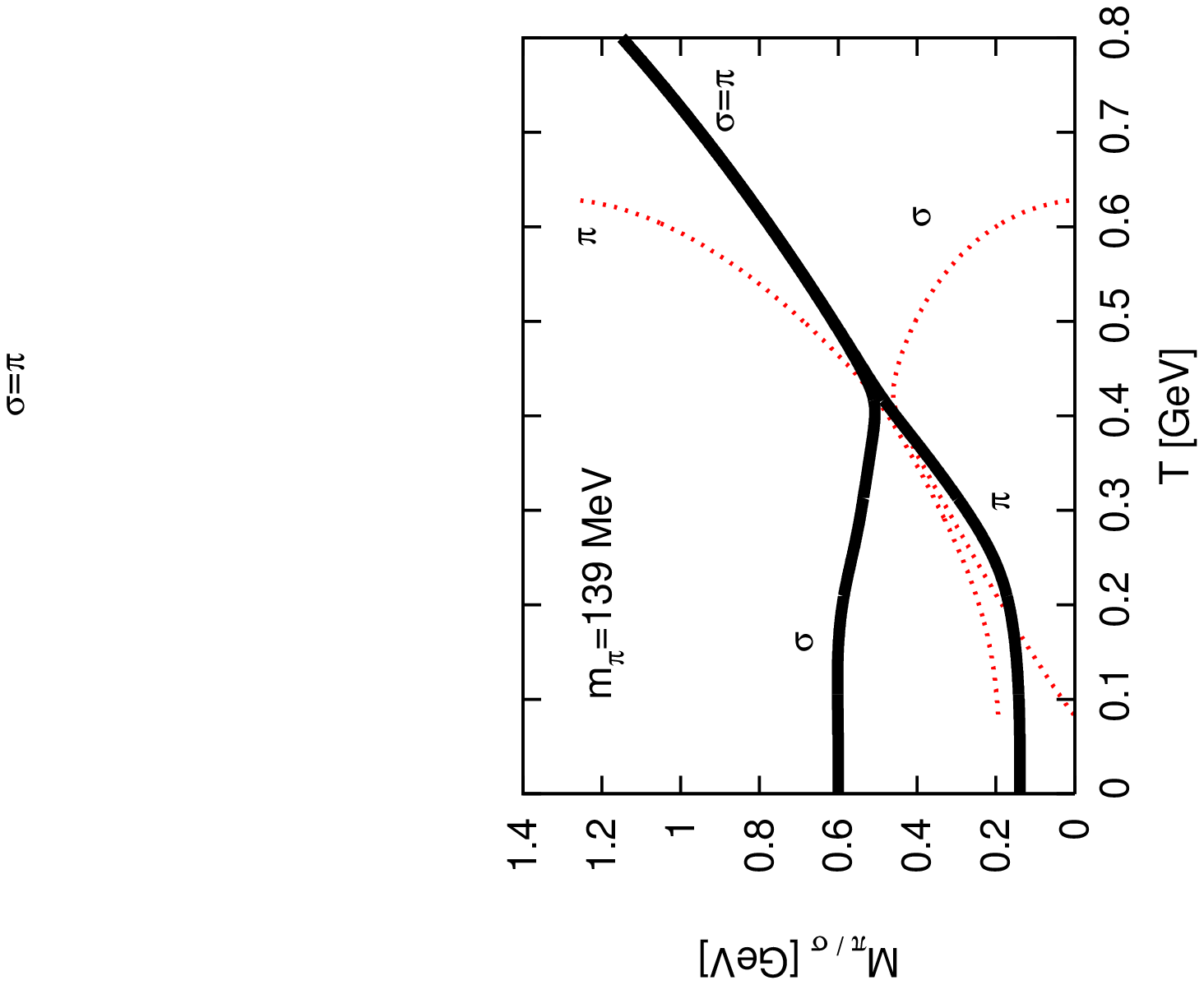}
\caption{Meson masses  as
  functions of temperature for $\mu=$ 600 MeV and $m_\pi=$ 139 MeV
  case. Stable branches are 
  presented by solid lines, whereas metastable ones -- by 
  dashed lines.} 
\label{fig-zm:mp139-600}
\end{figure}

\begin{figure}[ht]
\includegraphics[width=5.5cm,angle=-90]{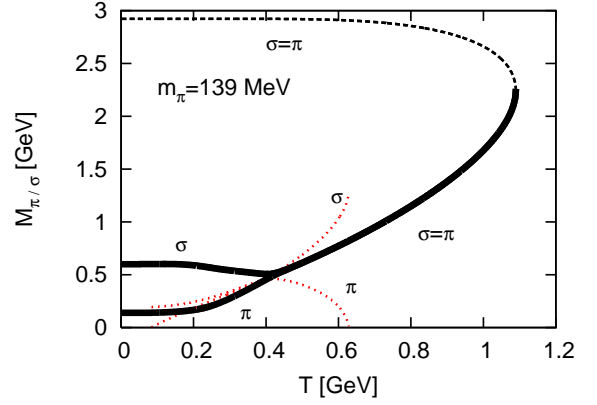}
\caption{Meson masses of Fig.~\ref{fig-zm:mp139-600} but in wider
  temperature region.}
\label{fig-m:mp139-600}
\end{figure}

\begin{figure}[ht]
\includegraphics[width=5.5cm,angle=-90]{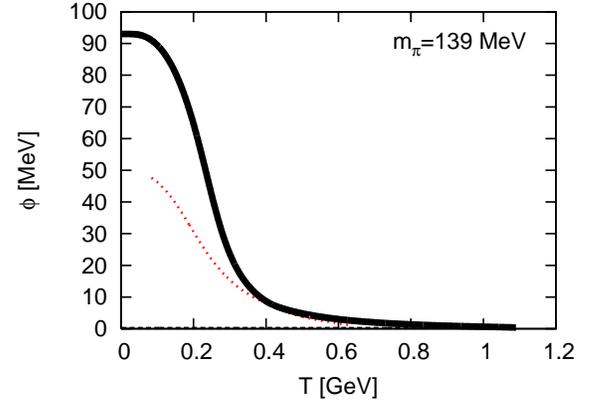}
\caption{The same as in Fig.~\ref{fig-zm:mp139-600} but for the field
  $\phi$.}
\label{fig-f:mp139-600}
\end{figure}

In the case of explicitly broken symmetry ($m_\pi=139$~MeV), the structure
of solutions at various $\mu$ scales is similar to that described above
for the chiral limit. Even the behavior of metastable branches
remains similar. We illustrate the changes on the example of $\mu=600$~MeV,
see Figs.~\ref{fig-zm:mp139-600}--\ref{fig-f:mp139-600} which looks the
most physically appealing and is close to results of the naive
renormalization \cite{IRK05}. The main difference from the $m_\pi=0$ case
is that the sequence of two phase transitions is transformed here into a
smooth cross-over transition.


\section{Conclusion}

We have studied a renormalized version of the gapless $\Phi$-derivable HF
approximation to the $\lambda\phi^4$ theory with spontaneous breaking of
the $O(N)$ symmetry, proposed in Ref.~\cite{IRK05}. This gHF approximation
simultaneously preserves all the desirable features of $\Phi$-derivable
approximation schemes (i.e., the validity of conservation laws and
thermodynamic consistency) and respects the NG theorem in the phase of
spontaneously broken symmetry. This is achieved by adding a correction
$\Delta\Phi$ to the conventional $\Phi$ functional.  The nature of this
correction can be understood as follows. The conventional $\Phi$-derivable
HF approximation cuts off an infinite series of diagrams, among which are
those providing the NG theorem in the phase of spontaneously broken
symmetry. By introducing the $\Delta\Phi$ correction to the HF
approximation we take into account a part of those omitted diagrams (at the
level of the actual approximation), which restores the NG theorem.

An advantage of the gHF approximation is that it allows for
scale-independent renormalization in the vacuum, unlike the conventional HF
approximation~\cite{Lenaghan}. The scale independence in the vacuum is a
direct consequence of the NG theorem, which makes the equations for the
pion self-energy and classical field degenerated.  However, even in the gHF
approximation, only renormalization scales higher than a certain value
($\mu_0$, cf. Eq. (\ref{mu-0})) are allowed, in order to ensure stability
of the renormalized approximation.

Nevertheless, the scale dependence still persists at finite temperatures.
The violation of renormalization-scale independence of $\Phi$-derivable
approximations was shown in~\cite{BP02,Destri05} from the point of view of
the renormalization-group $\beta$ function. There the $\beta$ function,
evaluated from the $\Phi$-functional formalism, was shown to deviate from
its perturbative expansion, beginning at orders in the expansion parameter,
higher than that explicitly taken in the $\Phi$ functional. The reason is
the violation of ``crossing symmetry'' in the sense of~\cite{HK1}: Solving
the self-consistent equations of motion corresponds to a partial
resummation of diagrams to any order in the expansion parameter (e.g., the
coupling constant $\lambda$ or $\hbar$, i.e., the order of loops in
perturbative Feynman diagrams) which is necessarily incomplete for any
truncation of the $\Phi$ functional.

Within our renormalization scheme, it becomes clear that the
renormalization-scale dependence at finite temperatures originates from the
subtraction of the ``hidden subdivergence'' of the four-point function
inside the self-consistent tadpole loop. As shown in~\cite{HK1}, this
four-point function consists of a resummation in only one channel, and thus
the $\beta$ function of this resummed four-point function deviates from the
correct one at orders higher than contained in the approximation of the
$\Phi$-functional, i.e., to $\mathcal{O}(\lambda^2)$.

At large scales ($\mu\gsim m_\sigma$) the chiral phase transition proceeds
similar to that in the naive renormalization scheme \cite{IRK05}. In the
case of the exact $O(N)$ symmetry, it proceeds through a sequence of two
second-order phase transitions rather than a single one.  In the first
transition the mean field vanishes but the meson masses still remain
different. The temperature of this phase transition, $T_2\simeq 180$~MeV,
is approximately $\mu$-independent and has approximately the same value as
that in the naive renormalization scheme \cite{IRK05}. In the second
transition also the masses become equal, and the $O(N)$ symmetry is
completely restored. The corresponding temperature, $T_{\scr{cross}}$,
turns out to be essentially $\mu$-dependent. When the $O(N)$ symmetry is
explicitly violated, the sequence of two phase transitions is transformed
into a smooth cross-over transition. Moreover, at $\mu\simeq m_\sigma$ the
results are even qualitatively close to those obtained with the naive
renormalization \cite{IRK05}. At small scales (say $\mu\lsim m_\sigma/2$),
the phase structure becomes very complicated, however still respecting the
NG theorem in the phase of spontaneously broken symmetry.

Another result concerns both the conventional renormalized HF and gHF
approximations, which in fact are identical in the phase of restored $O(N)$
symmetry. There exists an end point of the solution, i.e., a temperature
$T_{\scr{end}}$ above which there are no solutions to the gap equations.
The occurrence of such an end point in the HF approximation was first
pointed out by Baym and Grinstein~\cite{Baym-Grin} and is caused by the
dominant role of $\ln(M^2/\mu^2)$ terms in the gap equations, which
originate from the renormalization procedure.  The dominant behavior of the
$\ln(M_{\pi/\sigma}^2/\mu^2)$ terms signals a breakdown of the HF
approximation and the need to include higher-order corrections into the
$\Phi$ functional~\cite{Baym-Grin}. Another interesting question is whether
one can find a renormalization-group improved $\Phi$-derivable
approximation to cure this problem. We have found that at low scales (like
$\mu\lsim m_\sigma/2$), the end point appears at rather low temperatures,
leaving almost no room for the HF (as well as gHF) approximation in the
phase of restored $O(N)$ symmetry. At the same time, at ($\mu\gsim
m_\sigma$), the end point moves to rather high energies $T_{\scr{end}}\gsim
1.2$~GeV, hence allowing this approximation at least at $T\lsim 1$~GeV.

Summarizing, we have found that the gHF approximation is certainly
advantageous as compared to the conventional HF one, since it respects the
NG theorem and allows scale-independent renormalization in the vacuum.
These properties are closely interrelated. They both require that the set
of equations of motion are degenerate in the phase of spontaneously broken
symmetry. Nevertheless, there still are serious problems with the
renormalization of $\Phi$-derivable approximations for a theory with a
spontaneously broken symmetry. At finite temperatures the predictions of
renormalized gHF approximations essentially depend on the renormalization
scale, contrary to the case of the renormalized perturbation theory. In
this respect, the gHF approximation becomes similar to the leading-order
$1/N$-approximation, where also both the NG theorem and scale-independent
renormalization in the vacuum hold true~\cite{Lenaghan}. However, in view
of the medium-independent renormalization performed in accordance with
Refs.~\cite{HK3,HK1,Reinosa1}, this scale-dependence at finite temperatures
cannot be already interpreted as an artifact of temperature-dependent
counter terms. It may turn out that this scale dependence is a consequence
of triviality of the $\lambda\phi^4$ theory~\cite{Consoli} which, when it
is renormalized, therefore requires an external scale to serve for a scale,
below which it can be used as an effective field theory to describe the
low-energy phenomenology.

\acknowledgments

We are grateful to B. Friman and D.N. Voskresensky for useful discussions.
One of the authors (Y.I.) acknowledges partial support by the Deutsche
Forschungsgemeinschaft (DFG project 436 RUS 113/558/0-2), the Russian
Foundation for Basic Research (RFBR grant 03-02-04008) and Russian
Minpromnauki (grant NS-1885.2003.2). H.v.H. acknowledges partial support by
the U.S. National Science Foundation under grant PHY-0449489 and by the
Alexander von Humboldt Foundation as a Feodor-Lynen fellow.


\end{fmffile}
\end{document}